\documentclass[aps,prl,epsfig,twocolumn,showpacs,superscriptaddress]{revtex4}
\usepackage{amsmath}
\usepackage{epsfig}
\usepackage{graphics,color}

\begin{document}
\title{Twin peaks in rf spectra of Fermi gases at unitarity}

\author{P.~Massignan}
\affiliation{\mbox{Institute for Theoretical Physics, Utrecht University, Leuvenlaan 4, 3584 CE Utrecht, The Netherlands.}}
\affiliation{Laboratoire Kastler Brossel, \'Ecole Normale Sup\'erieure, 24 rue Lhomond, 75005 Paris, France}
\author{G.~M.~Bruun}
\affiliation{ Niels Bohr Institute, Universitetsparken 5, DK-2100 Copenhagen \O, Denmark.}
\author{H.~T.~C. Stoof}
\affiliation{\mbox{Institute for Theoretical Physics, Utrecht University, Leuvenlaan 4, 3584 CE Utrecht, The Netherlands.}}

\date{\today}

\begin{abstract}
 We calculate the radio-frequency spectrum of balanced and imbalanced ultracold Fermi gases in the normal phase at unitarity.
 For the homogeneous case the spectrum of both the majority and minority components always has a single peak even in the pseudogap regime.
 We furthermore show how the double-peak structures observed in recent experiments arise due to the inhomogeneity of the trapped gas.
 The main experimental features observed above the critical temperature in the recent experiment of Schunck \emph{et al.} [Science \textbf{316}, 867, (2007)] are recovered with no fitting parameters.
\end{abstract}

\maketitle
 Pairing and superfluidity play a central role in the physics of interacting Fermi gases spanning a wide range of energies from neutron stars and quark-gluon plasma over conventional superconductors to cold atomic gases.
 For weakly-interacting systems the two phenomena appear simultaneously, whereas for strongly correlated systems pairing sets in at temperatures $T^*$ higher than the critical temperature for superfluidity $T_c$.
 This raises intriguing questions concerning the nature of the paired state for a strongly-interacting system. 
 Radio-frequency (rf) experiments on cold atomic gases at unitarity enable the study of such fundamental questions~\cite{Regal,Chin}.
 In particular, recent experiments where superfluidity is quenched by a large population imbalance examine the strongly correlated normal state~\cite{Schunck,Shin}.
 At low temperatures, the shifted single peak spectrum was attributed to the presence of pairing without superfluidity.
 For intermediate temperatures, a double-peak structure of the rf spectrum was observed.
It was speculated whether this double-peak comes from the local co-existence of paired and unpaired atoms as in the BEC regime.
For a balanced system at very low temperatures, in-situ measurements indicate that no such double-peak structure exists for a homogeneous system in the superfluid phase~\cite{Shin}.

A theoretical description of a strongly-interacting Fermi gas is challenging, in particular at nonzero temperatures~\cite{Giorgini}.
 In the present paper, 
 we present such an analysis based on a diagrammatic approach that includes finite temperature pair correlations in the normal phase.
 At the single-particle level, the theory is shown to compare favorably with Monte-Carlo calculations in the limit of a very large population imbalance. 
We find that for a homogeneous system in the normal phase the rf spectrum is characterized at all temperatures $T$ and imbalances by a single peak, even in the pseudogap regime where strong interactions suppress the single-particle density of states around the Fermi energy.
 This holds for \emph{both} the majority and minority components indicating that a simplified picture of unpaired atoms co-existing with non-condensed pairs is too simple at unitarity. 
  Interaction effects become more important with decreasing $T$, and the spectra shift to higher energies as compared to the bare atomic resonance. 
 For trapped systems, we demonstrate that the presence of double peaks in the spectrum is due to the inhomogeneity of the sample.
 In the large imbalance case, we recover the main features of the rf spectrum observed 
  by the MIT group~\cite{Schunck}. 

We consider a gas of fermionic atoms with mass $m$ in a potential $V({\mathbf{r}})=m(\omega_x^2x^2+\omega_y^2y^2+\omega_z^2z^2)/2$. Initially, 
the atoms are prepared in a mixture of two hyperfine states $|1\rangle$ and $|2\rangle$ with an interaction characterized by the $s$-wave scattering 
length $a_{12}$. An essentially uniform radio-frequency beam couples the state $|2\rangle$ to a third hyperfine state $|3\rangle$ which initially is empty. We describe the effect of the rf beam on the atoms with the operator $(\hbar=1)$
\begin{equation}
H_{\rm  rf}=\frac{\Omega}{2}\int d\mathbf{r}\left[e^{-i\omega t}\psi_3^\dagger({\mathbf{r}},t)\psi_2({\mathbf{r}},t)+\rm{h.c.} \right],
 \label{RFTerm}
\end{equation}
where $\psi_i({\mathbf{r}},t)$ is the field operator for the atoms in state $|i\rangle$, $\Omega$ is the Rabi frequency describing 
the coupling of the involved hyperfine states to the electromagnetic field, and $\omega$ is the rf frequency. The induced transition rate $R(\omega)$ from state $|2\rangle$  to $|3\rangle$ is  within linear response given by
\begin{equation}
R(\omega)\propto-{\rm Im}{\mathcal{D}}(\omega) \equiv -\int d\mathbf{r}d\mathbf{r}'{\rm Im}{\mathcal{D}}({\mathbf{r}},{\mathbf{r}}',\omega)
\label{Linresp}
\end{equation}
where ${\mathcal{D}}({\mathbf{r}},{\mathbf{r}}',\omega)$ is the Fourier transform of the retarded pseudospin flip correlation function
$-i\theta(t-t')\langle[\psi_3^\dagger({\mathbf{r}},t)\psi_2({\mathbf{r}},t),\psi_2^\dagger ({\mathbf{r}}',t')\psi_3({\mathbf{r}}',t')]\rangle$.  

\begin{figure}
\includegraphics[width=0.8\columnwidth,height=0.5\columnwidth,clip=]{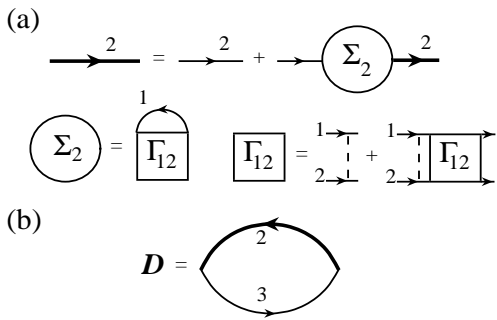}
\caption{(a) The atomic propagator for the  $|2\rangle$ atoms: thin (thick) lines are non-interacting (interacting) Green's functions, $\Sigma$ is the self-energy and $\Gamma$ is the T-matrix treating the interaction with the $|1\rangle$ atoms in the ladder approximation.
(b) The correlation function ${\mathcal{D}}$.}
\label{FeynFig}
\end{figure}

For the two hyperfine states of  $^6$Li used in the MIT experiment, a broad Feshbach resonance 
with  $|a_{12}|\rightarrow\infty$ is located at $834$G.
Close to resonance there are  strong correlations between the atoms 
in states $|1\rangle$ and $|2\rangle$. Furthermore, the state $|3\rangle$ 
in general interacts with the state $|1\rangle$~\cite{23inter}.
 This makes a consistent calculation of the correlation function ${\mathcal{D}}$ a complicated problem. 
In the present paper, we use a many-body theory which includes the 
two-particle physics leading to the 
Feshbach resonance. This corresponds to treating  the interaction between 
 $|1\rangle$ and $|2\rangle$ atoms in the ladder approximation as discussed extensively in the literature~\cite{Randeria}. 
The diagrammatic structure of our calculational scheme is shown in Fig.~\ref{FeynFig}.
We neglect for simplicity the interaction between states $|1\rangle$ and $|3\rangle$.
The correlation function ${\mathcal{D}}$ then decomposes 
into a propagator of $|2\rangle$ atoms interacting with the $|1\rangle$ atoms and a free propagator of $|3\rangle$ atoms. 
Assuming that the population of state $|3\rangle$ remains negligible throughout the experiment,
 the transition rate (\ref{Linresp}) for a homogeneous system can be written as 
\begin{equation}
 {\rm Im}\mathcal{D}(\omega)=-{\mathcal{V}}
 \int\frac{d\mathbf{k}}{(2\pi)^3} A_2(k,\xi_{2k}-\omega)f(\xi_{2k}-\omega)
 \label{D}
\end{equation}
with ${\mathcal{V}}$ the volume of the system.
Here $\xi_{ik}=k^2/2m-\mu_i$ is the kinetic energy of the $|i\rangle$ atoms relative to their  
chemical potential $\mu_i$ and $f(\epsilon)=[\exp(\beta\epsilon)+1]^{-1}$. 
All propagators are calculated within the imaginary time formalism and Wick rotated 
 $i\omega_n\rightarrow \omega+i0_+$ to obtain the retarded correlation function 
 ${\mathcal{D}}(\omega)$ at the relevant rf frequency.   
 We measure frequencies relative to the $|2\rangle$-$|3\rangle$ resonance in the absence of atoms in state $|1\rangle$.
  The formalism is described in detail in Ref.~\cite{BruunBaym} and here extended to the case $\mu_1\ne\mu_2$. 
  This allows us to model experiments with population imbalance $\delta=(N_1-N_2)/(N_1+N_2)$, where $N_i$ is the number of $|i\rangle$ 
 atoms in the trap and $N_1 \geq N_2$.

We treat  the trapping potential in the Thomas-Fermi approximation calculating the rf spectrum for the trapped system by 
integrating  over the trap the homogeneous signal obtained from Eq.~(\ref{D})
with local chemical potentials $\mu_i({\mathbf{r}})=\mu_i-V({\mathbf{r}})$.
The correlation function ${\mathcal{D}}(\omega)$  obeys the sum rule
 \begin{equation}
 \int \frac{d\omega}{\pi}{\rm Im}\mathcal{D}(\omega)=N_2.
 \label{sumrule}
 \end{equation}
The global chemical potentials $\mu_1$ and $\mu_2$ must be adjusted such that they reproduce the 
correct $N_1$ and $N_2$. The density profiles of atoms $|1\rangle$ and $|2\rangle$ may be obtained from the spectral functions as 
$n_i({\mathbf{r}})=\int d\mathbf{k}\int d\omega A_i(k,\omega)f(\omega)/(2\pi)^{4}$. Also from this expression one may calculate $N_1$  and $N_2$.
 For large imbalance, it is a good approximation to take the ideal gas values for $n_1({\mathbf{r}})$ and $\mu_1$ since $N_1 \gg N_2$. 
 We have checked that determining $N_2$ from the interacting density profile and from Eq.~(\ref{sumrule}) yields consistent results. 
 
Consider first the rf spectrum of a homogeneous and balanced system at unitarity. 
The system is in the normal phase for  $T>T_c$, where $k_BT_c\approx0.26E_F$ 
from the Thouless criterion  with the thermodynamic potential calculated in the ladder approximation~\cite{BruunSmith,Perali2}. 
The Fermi energy is defined as $E_F=(6\pi^2n)^{2/3}/2m$, with $n=n_1=n_2$.
We determine the chemical potentials $\mu_1(T)=\mu_2(T)$ self-consistently in order to keep the density  $n$ fixed.

\begin{figure}
\includegraphics[width=0.8\columnwidth, height=0.60\columnwidth,clip=]{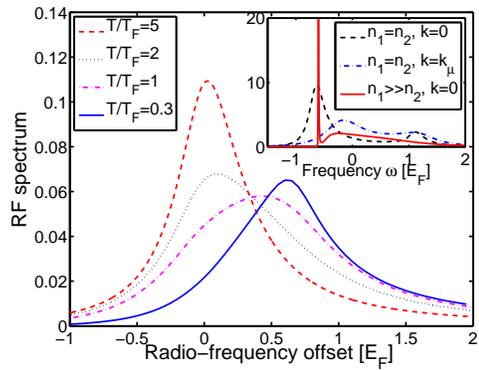}
\caption{(Color online) Rf spectrum of a homogeneous Fermi gas with equal populations. Inset: spectral function for a balanced system at $T=0.3T_F$ and for the minority component of an imbalanced system at $T=0.05T_F$. ($k_\mu=\sqrt{2m\mu}$).}
\label{homogFig}
\end{figure}

In the inset of Fig.\ \ref{homogFig} we plot the spectral function $A_2(k,\omega)=-2{\rm Im}[G_2(k,\omega)]$ at $T\gtrsim T_C$ where the system is in
 the pseudogap regime: at low momenta $A_2$ has the characteristic double-peak structure, leading to a suppression of the density 
of states around the Fermi energy~\cite{Perali,BruunBaym,Boldizsar}. 
To check the accuracy of our approach, we also plot in the inset of Fig.\ \ref{homogFig}
the spectral function $A_2$ in the case of a single $|2\rangle$ atom in a sea of $|1\rangle$ atoms for very low $T$.
  We see that there is a narrow quasiparticle peak at $\omega=-0.6E_F$, in full agreement with a recent variational quantum Monte Carlo calculation~\cite{Lobo}. 
  This indicates that the ladder approximation used in the present paper accounts in the extremely imbalanced case for most of the correlation energy at unitarity confirming the accuracy of our results~\cite{exchange}. 

\begin{figure}
\includegraphics[width=0.8\columnwidth, height=0.60\columnwidth,clip=]{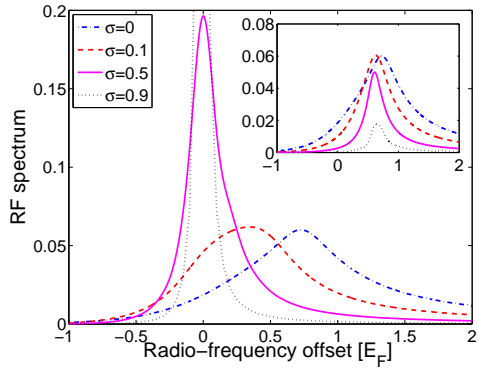}
\caption{(Color online) Rf spectrum of the majority and minority components (respectively, in the main figure and in the inset) of a homogeneous gas for various imbalances $\sigma=(n_1-n_2)/(n_1+n_2)$ at $T=0.3T_F$ and constant $n_1$.}
\label{MajorAndMinorFig}
\end{figure}

 In Fig.\ \ref{homogFig} we plot the rf spectrum as a function of temperature for the balanced case. 
 We see that the integration over momenta in Eq.~(\ref{D}) has washed out the double-peak structure of the spectral function, and the spectrum always consists of a single peak.
 In the classical regime, the $|1\rangle$-$|2\rangle$ cross section scales as $\sigma\propto \lambda_T^2$ with $\lambda_T=\sqrt{2\pi/m k_B T}$ the thermal deBroglie wavelength. 
 The scattering rate thus becomes suppressed with increasing temperature and the system approaches ideal gas behavior, exhibiting a narrow spectral line centered at the atomic frequency $\omega=0$.  
 With decreasing temperature the rf line smoothly shifts to higher energies due to increasing pairing interactions between $|1\rangle$ and $|2\rangle$ atoms, but \emph {no double-peaks} appear in the spectrum of a homogeneous system above $T_c$, even though the system is in the pseudogap regime with a double-peaked spectral function.
 Similar asymmetric spectra with suppressed weight at $\omega=0$ were observed for the balanced system at lower $T$ in the superfluid phase~\cite{Shin, Perali3}. 
 In Fig.~\ref{MajorAndMinorFig} we plot the spectrum as a function of imbalance for fixed temperature and density $n_1$ \cite{majorSpec}.
 The offset of the peak for one component is mainly determined by the density of the other component.
 The majority spectrum smoothly changes from a broad shifted peak to a narrow unshifted peak, in accord with the fact that the majority component at high imbalance is an ideal gas.
 The minority spectrum instead remains broad and its offset is roughly independent of imbalance. We have verified that this physical picture holds at all temperatures down to $T=0$ in the region where the gas is in the normal phase.
  This indicates that the picture of molecules mixed with unbound atoms no longer holds at unitarity where the pairing is rather a many-body phenomena.
 By comparing the width of the spectral peak at unitarity, which is of order $k_F^2/m$, with the molecular energy, of order $1/ma^2$, we expect the co-existence picture  to become valid on the BEC side of the resonance when $k_Fa\lesssim1$.

\begin{figure}
\includegraphics[width=0.8\columnwidth,height=0.60\columnwidth,clip=]{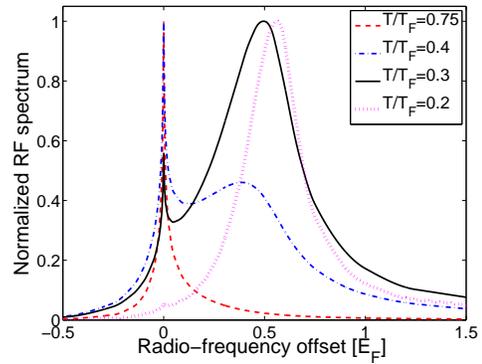}
\caption{(Color online) Rf spectrum of the minority component of a trapped Fermi gas with population imbalance $\delta=0.9$. Each spectrum satisfies the sum rule Eq.\ (\ref{sumrule}), but here curves are normalized such that they reach 1 at their highest peak.}
\label{trapFig}
\end{figure}

\begin{figure}
\includegraphics[width=0.8\columnwidth,height=0.60\columnwidth,clip=]{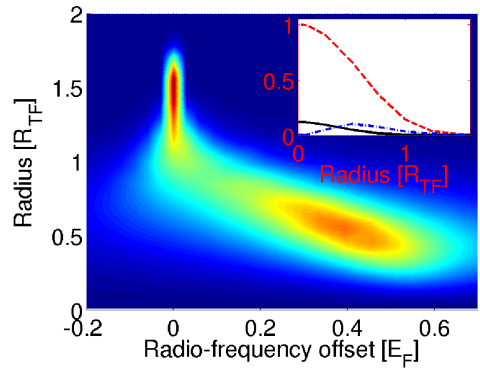}
\caption{(Color online) Surface plot of the integrand $r^2 {\rm Im}\mathcal{D}(r,\omega)$ of the trapping average at $\delta=0.9$ and $T=0.4T_F$ for the minority component. In the inset, we plot $n_1(r)/n_1(0)$ (dashed), $n_2(r)/n_1(0)$ (continuous) and $r^2 n_2(r)/R_{TF}^2 n_2(0)$ (dot-dashed).}
\label{integrandFig}
\end{figure}

We now consider the trapped case. In Fig.\ \ref{trapFig} we plot the rf spectrum of the minority component for different temperatures at 
imbalance $\delta=0.9$ corresponding to the experimental conditions of Figs.\ 2(a-d) in Ref.\ \cite{Schunck}. 
The Fermi temperature  is defined as $k_BT_F=(6N_1)^{1/3}\bar\omega$, with $\bar{\omega}^3=\omega_x\omega_y\omega_z$.
Within our theory, the gas does not become superfluid at this high imbalance.
 At high $T$ one recovers an ideal gas single peak centered at  $\omega=0$. The spectral line is broadened by pairing correlations, and its width is comparable to what is  experimentally observed, see Fig.\ 2(a) of Ref.\ \cite{Schunck}. 
As the temperature decreases, a broader peak emerges from the atomic one; one peak is narrow and centered  at $\omega=0$ for all $T$ whereas the other peak moves to higher frequencies with decreasing $T$, in agreement with the experiments~\cite{Chin,Schunck}.
 To analyse closer the origin of this double-peak structure, we plot in Fig.\ \ref{integrandFig} for $T=0.4T_F$ the integrand of the trapping average $r^2 {\rm Im}\mathcal{D}(r,\omega)$,
where with $\mathcal{D}(r,\omega)$ we mean the correlation function evaluated with the local chemical potentials $\mu_i(r)$
($\bar{\omega}^2r^2=\omega_x^2x^2+\omega_y^2y^2+\omega_z^2z^2$). 
One sees that for any given radius [corresponding to a homogeneous system with densities $n_1(r)$ and $n_2(r)$], there is always a single peak in the rf spectrum in agreement with the conclusions obtained above for a homogeneous system. 
For large $r$ the gas is dilute and the 
spectrum displays locally a sharp peak at the bare atomic frequency.
 This gives rise to the $\omega=0$ peak in the rf spectrum for the trapped gas. 
As $r$ decreases towards the center of the cloud, the rf peak shifts continuously towards higher energies as the pair 
correlations between species $|1\rangle$ and $|2\rangle$ increase with increasing densities $n_1(r)$ and $n_2(r)$. 
The integrand contributes most to the total rf signal for radii corresponding to the highest
value of $r^2n_2(r)$; as shown in the inset of Fig.\ \ref{integrandFig} this corresponds at $T=0.4T_F$ to $r/R_{TF}\approx0.5$
with $R_{TF}=k_F/m\bar\omega$ the  Thomas-Fermi radius of the majority cloud. For this value of $r$,  the rf peak 
is located at $\omega\approx0.5E_F$. 
This is the origin of the second peak located at $\omega\approx0.5E_F$ for the trapped system at $T=0.4T_F$.
As the gas gets colder, $\mathrm{max}_{r} [r^2n_2(r)]$ shifts towards $r=0$ and the pair correlations increase leading to the 
pairing peak moving towards higher frequency.
 By varying the imbalance $\delta$, we find that the peak position is roughly independent of the imbalance, in agreement with experiments~\cite{deltaInvariance} and with our findings for a homogeneous gas (see Fig.~\ref{MajorAndMinorFig}).
 At very low T the atomic peak disappears: the whole cloud is now in the pseudo\-gap regime and the spectrum consists of a single broad and shifted peak.
Note that our theory obtains a remarkable agreement with the experimental findings without any fitting parameter.
It also reproduces reasonably well the experimental 
values for the shift and temperature dependence of the pairing peak although 
the peak is predicted to emerge at a  somewhat lower temperature and  with a larger shift 
than what is  observed by the MIT group. 
 This disagreement could be due to the fact that we have neglected interactions between states $|1\rangle$ and $|3\rangle$, which are significant at the considered magnetic field. To calculate the effects of the  $|1\rangle$-$|3\rangle$ interactions requires the inclusion of a vertex correction in the correlation function ${\mathcal{D}}$, the 
so-called Aslamazov-Larkin contribution~\cite{AL}. On general grounds, we expect the inclusion of the  $|1\rangle-|3\rangle$ interactions 
to reduce the spectral width and the shift of the pairing peak. Sum rule arguments show that this shift scales as $a_{12}^{-1}-a_{13}^{-1}$ for interactions 
with the same high energy behavior, and therefore vanishes in the limit $a_{12}=a_{13}$~\cite{Baym,Punk}. 
The inclusion of the $|1\rangle$-$|3\rangle$ interaction in the superfluid phase was recently demonstrated to reduce the pairing peak shift~\cite{Perali3}.


We briefly compare our analysis to other recent works. 
The results in \cite{He,Kinnunen04,KinnunenPRL} are based on a phenomenological theory designed to describe pseudogap physics, in which an unknown parameter gives rise to a suppression in the density of states above $T_C$.
Using this theory, it was argued that the double peaks observed in the rf spectrum of a balanced gas \cite{Chin} are due to the presence of the trap \cite{Kinnunen04}, in agreement with our conclusions.
 However, in \cite{KinnunenPRL} it was claimed that a double peak should also exist for a homogeneous system, at variance with our findings.
After the submission of our work, it was argued that the appearance of double peaks in the trapped spectrum can be traced back to the functional form of the equation of state for a fermionic gas~\cite{Mueller2}.
 The approach in \cite{Perali3} is similar to our analysis; they however focus on the $T=0$ homogeneous superfluid state.

In conclusion, our theory reproduces the main features of the observed rf spectrum for a trapped gas in the normal phase,
 i.e., the unshifted $\omega=0$ peak 
which disappears with lowering $T$, the emergence of the pairing peak leading to a double-peak structure for intermediate $T$,
the single pairing peak for low $T$, and the independence of the pairing peak position on the imbalance. We have also demonstrated how the recent rf experiments in combination with a proper microscopic theory provide insight into the 
nature of pairing for strongly-interacting imbalanced Fermi gases.

\acknowledgments{Part of this work was done while two of us (P.~M.~and G.~M.~B.) stayed at the Institut Henri Poincar\'e-Centre Emile Borel.
We thank this institution and the IFRAF for hospitality and support.
Useful discussions with M.\ Zwierlein and especially C.\ J.\ Pethick are acknowledged.}


\begin{thebibliography}{99}
\bibitem{23inter}For coherent rf excitations, we can neglect the 
 interactions between states $|3\rangle$  and $|2\rangle$.
\bibitem{exchange} The exchange diagrams involving creation of particle-hole pairs of species $|2\rangle$, which are not included in the ladder sum, are indeed suppressed in the limit $n_2 \ll n_1$.
\bibitem{majorSpec} The majority spectrum is obtained from Eq.~(\ref{D}) by interchanging the subscript 2 with 1.
\bibitem{deltaInvariance} Over the range $0.5<\delta<0.99$ we have observed peak position shifts of less than $15\%$.
\end{thebibliography}
\end{document}